# Design drivers for a wide-field multi-object spectrograph for the William Herschel Telescope

Marc Balcells[*a,b], Chris R. Benn[a], David Carter[c], Gavin B. Dalton[d,e], Scott C. Trager[f], Sofia Feltzing[i], Marc A. W. Verheijen[f], Matt Jarvis[g], Will Percival[h], Don C. Abrams[a], Tibor Agocs[a], Anthony G. A. Brown[k], Diego Cano[a], Chris Evans[l], Amina Helmi[f], Ian J. Lewis[d], Ross McLure[l], Reynier F. Peletier[f], Ismael Pérez-Fournon[b], Ray M. Sharples[m], Ian A. J. Tosh[e], Ignacio Trujillo[b], Nic Walton[j], Kyle B. Westfall[f]

[a]Isaac Newton Group of Telescopes, Apartado 321, 38700 Santa Cruz de La Palma, Canary Islands, Spain;
[b]Instituto de Astrofísica de Canarias, C/ Vía Láctea S/N, 38200 La Laguna, Tenerife, Spain, and Departamento de Astrofísica, Universidad de La Laguna, 38200 La Laguna, Spain;
[c]Astrophysics Research Institute, Liverpool John Moores University, Twelve Quays House, Egerton Wharf, Birkenhead, Wirral, CH41 1LD, UK;
[d]Physics Department, University of Oxford, Keble Road, Oxford, OX1 3RH, UK;
[e]STFC-Rutherford-Appleton Laboratory, HSIC, Didcot, OX11 0QX, UK;
[f]Kapteyn Astronomical Institute, University of Groningen, Postbus 800, 9700 AV Groningen, The Netherlands;
[g]Centre for Astrophysics Research, Science & Technology Research Institute, University of Hertfordshire, Hatfield, AL10 9AB, UK;
[h]Institute of Cosmology and Gravitation, University of Portsmouth, Portsmouth, PO1 3FX, UK;
[i]Lund Observatory, Box 43, SE-221 00 Lund, Sweden;
[j]Institute of Astronomy, University of Cambridge, Madingley Road, Cambridge CB3 0HA, UK;
[k]Leiden Observatory, Leiden University, P.O. Box 9513, NL-2300 RA Leiden, The Netherlands;
[l]UK Astronomy Technology Centre, Royal Observatory Edinburgh, Blackford Hill, Edinburgh, EH9 3HJ, UK;
[m]Centre for Advanced Instrumentation, University of Durham, South Road, Durham DH1 3LE, UK

**ABSTRACT**

Wide-field multi-object spectroscopy is a high priority for European astronomy over the next decade. Most 8-10m telescopes have a small field of view, making 4-m class telescopes a particularly attractive option for wide-field instruments. We present a science case and design drivers for a wide-field multi-object spectrograph (MOS) with integral field units for the 4.2-m William Herschel Telescope (WHT) on La Palma. The instrument intends to take advantage of a future prime-focus corrector and atmospheric-dispersion corrector (Agocs et al, this conf.) that will deliver a field of view 2 deg in diameter, with good throughput from 370 to 1,000 nm. The science programs cluster into three groups needing three different resolving powers R: (1) high-precision radial-velocities for Gaia-related Milky Way dynamics, cosmological redshift surveys, and galaxy evolution studies (R = 5,000), (2) galaxy disk velocity dispersions (R = 10,000) and (3) high-precision stellar element abundances for Milky Way archaeology (R = 20,000). The multiplex requirements of the different science cases range from a few hundred to a few thousand, and a range of fibre-positioner technologies are considered. Several options for the spectrograph are discussed, building in part on published design studies for E-ELT spectrographs. Indeed, a WHT MOS will not only efficiently deliver data for exploitation of important imaging surveys planned for the coming decade, but will also serve as a test-bed to optimize the design of MOS instruments for the future E-ELT.

**Keywords:** Spectrograph, multi-object, science requirements, optical design

---

[*] balcells@ing.iac.es

# 1 INTRODUCTION

It is clear that 10-m class telescopes currently lack, and will lack for the foreseeable future, an important capability – wide-field spectroscopy. With the exception of the Subaru telescope, large telescopes have not developed prime focus (PF) units; hence their maximum field of view (FOV) is a few tens of arcmin. Such FOV are suitable for deep, pencil-beam surveys, for example the DEEP surveys with Keck/DEIMOS[†], and the VVDS [41] and zCOSMOS [44] surveys with VLT/VIMOS. But such FOVs are too small for surveys covering hundreds or thousands of square degrees.

Wide-field multi-object spectroscopy will be in particular demand in the coming decade, given the many ambitious imaging surveys on the way, including Gaia [55], optical and NIR surveys with VISTA and VST [2], Pan-STARRS [36], UKIDSS [40], and radio surveys such as LOFAR [37]. The science goals of these surveys range over stellar and sub-stellar astronomy, Galactic astronomy, galaxy evolution and cosmology. In all cases, extracting physics from the catalogued sources requires spectroscopy. For both extra-galactic studies and studies of the Milky Way, complementary spectroscopic observations are crucial for full scientific exploitation. For extra-galactic studies, spectra provide redshifts, velocity dispersions and stellar population indicators; for Gaia, we need to complement the position and proper motion measurements for the fainter portion of the Gaia stars with radial velocities in order to obtain full 6D phase-space information of the dynamical state of the Galaxy.

All of these surveys require instruments with a large FOV. Furthermore, the magnitude range 15<V<21 that is appropriate to 4m apertures provides a good match to the continuum brightness distributions of many wide-field surveys, both Galactic and extra-galactic, as well as to sources selected to have strong emission lines in the optical (see Section 3.3). For these types of surveys, we argue that 4m class telescopes, equipped with wide-field capabilities, provide by far the most efficient method of obtaining redshifts and detailed spectral information.

Because of the importance of the Gaia mission, recent reviews of European astronomy such as the *Astronet Roadmap* and the Astronet-sponsored report of the *European Telescope Strategy Review Committee* (ETSRC)[‡] have put high priority on European astronomers gaining access to multi-object spectrographs on 4-m telescopes from both hemispheres. These reports put similar priority on allowing European astronomers to compete in the cosmological quest to constrain the equation of state of dark energy through massive redshift surveys of intermediate-redshift galaxies. Because of this, there are opportunities for the instrument to become a European facility.

In this paper, we lay out the science case and design drivers for a wide-field multi-object spectrograph for the 4.2-m William Herschel Telescope (WHT). The WHT is operated by the Isaac Newton Group of Telescopes (ING) on the Observatorio Roque de los Muchachos on La Palma, Canary Islands.

In Sections 3 and 4 below, we set out the science case for wide-field multi-object spectroscopy on a 4-m telescope, and summarise the top-level science requirements. In Sections 5 through 9, we discuss instrumental parameters and concepts that would deliver the required capabilities, as well as design options for the front end (multi-object, integral-field units) and the spectrograph. Design trade-offs are discussed in Section 10.

# 2 EFFORTS ELSEWHERE

A number of projects are underway to address similar science, mainly the cosmology aspects. Most of the projects address surveys for cosmology. They are at various levels of development and funding, from fully funded to abandoned.

- SDSS-III spectrographs – BOSS survey [60].
- LAMOST spectrograph.
- PRIMUS at Magellan: FOV 25' 5000 slits at R=40.
- AAT/Hermes: high-resolution version of AAOmega, using 2dF positioner. R=30,000.
- CAHA XMS 4000 slits R=400, slits 1.5"x10".
- BigBOSS spectrograph at Mayall.

---

[†] http://deep.ucolick.org/

[‡] Both ASTRONET reports are available at http://www.astronet-eu.org/

- WFMOS at Gemini, @Subaru [46][26]
- GYES at CFHT: FOV>1 degree, ~500 fibres, R=25,000.

# 3 SCIENCE CASE

The recent ETSRC report has analyzed the observational capabilities needed in the coming decade that can be provided on European intermediate-aperture telescopes. As already pointed out in Sec. 1, the emphasis is naturally on wide-field astronomy. The ETSRC recommendations fit well with the main science interests of the current ING community, namely massive multiplex multi-object spectroscopy and wide-field integral-field spectroscopy. In this section we develop the main science cases.

## 3.1 Milky Way archaeology: exploiting Gaia's scientific legacy

Our own Galaxy, the Milky Way, is the only galaxy for which we will be able in the foreseeable future to determine the precise chemo-dynamical formation and evolutionary history. The Gaia astrometric satellite [55] will revolutionize the study of the Milky Way and its companions, delivering photometry, positions and proper motions for every object on the sky down to V=20 mag—one billion including stars, galaxies, quasars, and solar system objects—within five years of its launch in 2012. The Gaia satellite also carries the Radial Velocity Spectrograph (RVS) which will obtain spectra centred on the Ca II IR triplet (845-870 nm) for all stars down to V=17. From these spectra, radial velocities as well as stellar parameters, e.g. overall metallicity, effective temperature, will be derived. Expected accuracies are, for radial velocities about 1 km/s for stars brighter than 13, worsening to an accuracy of 15 km/s at V=17[§]. In order to obtain the full 6D-phase space information for a large fraction of the stars observed by Gaia, complementary radial velocities are needed for fainter stars.

From the RVS spectra some information about stellar parameters can be derived. Here it is expected that the level of completeness will be significantly shallower than for the radial velocities. Hence, while (some) *dynamical* information will be available for stars down to V<17 mag from Gaia itself, information about the *formation times* of the stars—traced by their chemical compositions—will only be available for the brightest stars (and then only crudely).

For these reasons, Feltzing et al. (2009) have proposed a three-pronged strategy to complete Gaia's Galactic census[**]:

a. R=5000 multi-object spectroscopy to complete the full 6D phase-space information for stars fainter than V=17
b. R=20000 multi-object spectroscopy of metal-poor thick disk and halo stars (giants and dwarfs)
c. R=40000-60000 spectroscopy of selected metal rich (> 1 dex) stars in the disk and the (outer) halo.

The current instrument design drivers from Milky Way archaeology follow from the first two prongs of this strategy. Below we describe the major science drivers for these two prongs.

### 3.1.1 *The structure and history of the Galactic stellar halo and Galactic disk*

Understanding the mass content, the current dynamical state and the history of the Milky Way requires full 6D phase-space information for as many stars as possible over as large a volume as possible. For example, a clear prediction of cold dark matter (CDM) models of structure and galaxy formation is the existence of a large number of bound substructures in the halos of galaxies, orders of magnitude more abundant than the current population of satellites with stellar content observed in the Milky Way. It is possible that these structures are completely dark and will only be detected by their perturbations of the dynamics of visible structures, e.g., very cold stellar streams resulting from the accretion of satellites by the Milky Way [21][22]. Gaia will measure distances and proper motions for many such streams, but radial velocities for stars in these streams will be of insufficient accuracy (typical velocity dispersions in the streams are <5 km/s) or lacking. For streams that are even farther away, parallax errors will degrade the accuracy of tangential velocities, and radial velocities will be required. In the Galactic thin disk, we expect to uncover a large number of dynamically-induced substructures; unraveling their origins, whether resulting from internal evolution from spiral arms or the bar (or both), dispersion of clusters or even accretion of satellites, will require accurate radial velocities to obtain the full 6D phase space information necessary to unravel their history [6][17][18][12]. The velocity dispersion in the thin disk is ~10—30 km/s, requiring velocities accurate to <5 km/s to retrace the orbital histories of stars—a condition met by the tangential velocities but not (for the majority of stars throughout the thin disk) the radial velocities provided by Gaia/RVS. A kinematic map of the Galactic thick disk would allow for a determination of the orbital eccentricity as a function of galactocentric radius, a powerful diagnostic of mechanisms for its formation [57]. For all of these reasons, a survey of >$10^6$ stars over the entire sky with radial velocity accuracy of <5 km/s—but preferably 1—2 km/s—in the range 17<V<20 mag is crucial for completing Gaia's census.

---

[§] See the Gaia performance page at http://www.rssd.esa.int/index.php?project=GAIA&page=Science_Performance

[**] See the GREAT chemo-dynamical survey pages at http://camd08.ast.cam.ac.uk/Greatwiki/GreatCds

### *3.1.2 The merger history of the Milky Way through chemical labelling*

Another prediction of CDM galaxy formation models is that the Galaxy's stellar halo was built up by the accretion of many satellite galaxies (an idea with its origins at least as far back as Searle & Zinn 1978). The Milky Way currently has a fair number of faint companions. How are field halo stars related to these dwarf satellites? How much of the stellar halo was accreted, how are these accreted stars distributed throughout the halo, and when did these accretions occur? Solving these questions requires both dynamical *and* chemical information, linking the positions and velocities of stars with their nucleosynthetic histories. For example, it is becoming clear that the halo is not one monolithic structure but is in fact (at least) two halos, an inner and an outer halo, possibly with different formation histories [14][51].

Furthermore, several streams and overdensities have been uncovered in the outer halo by wide field photometric surveys (e.g. [45][7]), indicating that accretion has been important at least in the outer Galaxy.

Given the large number of streams and the possibility of "chemical labelling" of their members, and given the low density of halo stars, a survey that targets ~50,000 halo giants (expected to be in ~500 streams) to measure elemental abundances with accuracies of better than 0.1 dex/element would be sufficient to characterize the "building blocks" of our halo. Such a survey requires an area of ~2500 deg$^2$ at V<17—18 mag at R=20000. Chemical labelling is to be distinguished from "chemical tagging" which requires higher internal precision in the derived elemental abundances than can be achieved with R=20,000.

## 3.2 Probing the nature of dark energy

Understanding the current acceleration of the Universe is one of the most interesting problems facing physics today. Large surveys of galaxy redshifts provide two key methods by which we can measure the properties of dark energy. The galaxy clustering signal, and particularly features within it called Baryon Acoustic Oscillations act as a standard ruler, giving a tool with which to measure the geometry of the Universe as a function of redshift. Redshift-space distortions, apparent anisotropic patterns of galaxies, provide complementary information about the rate of build-up of large-scale structure. These distortions are caused by coherent comoving galaxy velocities, which lead to a measurable anisotropic clustering signal when these velocities are misinterpreted as being due to the Hubble flow. In addition, by comparing the strength of clustering for galaxies of different types, and by combining with CMB observations, a large galaxy survey will constrain galaxy formation models, the cosmological average matter density, the way galaxies trace the mass, the sum of the neutrino masses (which imprint a signal in the clustering pattern) and the primordial power spectrum shape created by inflation.

In order to make the next significant leap forwards in galaxy redshift surveys, we will need to provide of order $10^7$ spectra over at least 10,000 square degrees of the extragalactic sky. Such a survey would represent an order-of-magnitude improvement over the ongoing Baryon Oscillation Spectroscopic Survey (BOSS [60]), part of SDSS III. The survey will need to observe galaxies at redshifts between 0.6 and 1.4, selected from deep imaging surveys. To carry out a survey such as this in a reasonable total survey time, with ~1 hour integrations, even with a dedicated telescope, will require an instrument with a very wide field of view and multiplex capability. In order to get redshifts for the targeted emission line galaxies, undertake galaxy evolution science and Lyα cosmological studies, the desired wavelength coverage is from approximately 350 – 1100nm. A spectral resolution of 1000 to 2000 is adequate for redshifts. Multiplex needs to be as high as possible: values of 1000 lead to survey times of 10 years if bright nights are not useable.

## 3.3 Understanding galaxy formation and evolution

While the study of galaxy evolution is sometimes believed to be entirely a domain for 10-m class telescopes, many important areas are more efficiently explored with intermediate-size telescopes equipped with wide-field instruments, due to the combination of source densities on the sky and flux levels (either in continuum or in emission lines). In the following sub-sections we highlight a few examples. First, the full scientific exploitation of wide-field photometric surveys at optical, infrared or radio wavelengths will be curtailed unless optical spectroscopy for the sources is provided. A 4-m MOS therefore becomes essential to the success of key ground-based or space projects that required big investments. Second, investigation of the most recent phases of galaxy evolution, during which internal and environmental processes gave galaxies their present-day form, require wide fields; pencil-beam surveys on 10-m class telescopes sample small volumes at low redshifts. Third, in the local Universe, galaxy disks, which 30 years ago provided one of the first pieces of evidence for dark matter, remain today unique probes of the still unsolved distribution of luminous and dark matter on scales of tens of kpc. Their study is ideally tailored to a MOS spectrograph fitted with a monolithic IFU front end.

### 3.3.1 *The star formation history of the Universe: complementing radio and FIR probes*

Surveys at radio wavelengths will soon make a huge leap forward with the commissioning of LOFAR and the installation of the APERTIF receivers on the WSRT. LOFAR and APERTIF possibly offer the best way of tracing the star-formation history of the Universe without the principal bias inherent in optical surveys: dust obscuration. Deep radio surveys are dominated by star-forming galaxies (e.g. [61][71]), and due to the tight correlation between radio and far-infrared luminosity (e.g. [35]) radio surveys can be used as unbiased tracers of star-formation rate density of the Universe, analogous to the work carried out with HST in the mid-1990s [47]. Although much can be achieved using approximate photometric redshifts, these can only be obtained for radio sources with optical/near-IR counterparts, implying a similar dust-obscuration bias to that affecting past work with samples selected in the optical or near IR. Blind spectroscopy at the position of the radio sources, which should be feasible given the ~3 arcsec beam of LOFAR at 200 MHz, can provide emission-line redshifts for all star-forming galaxies at $z < 1.3$ using H$\alpha$ and [OII] emission lines and at $z > 2$ using Lyman-$\alpha$. As well as providing redshifts, the spectra can also be used to measure metallicity, and the stellar velocity dispersion, providing important information about the dynamical mass of the systems, and how this evolves with time.

Surveys with APERTIF will yield the spatially resolved distribution and kinematics of cold HI gas in $10^5$ galaxies in the nearby universe, providing redshifts, dynamical masses and (specific) angular momenta of gas-rich galaxies in various environments. APERTIF will also detect the radio continuum emission from $10^7$ galaxies at 1.4 GHz, corresponding to the same population of star-forming galaxies that LOFAR will detect. LOFAR will also be able to at least detect every active galactic nucleus (AGN), whether radio-loud or radio-quiet, out to very high redshifts. Many of these will be unresolved by LOFAR and thus will be difficult to distinguish from the star-forming galaxies. Only emission-line diagnostics from spectroscopy will allow us to determine the evolution of AGN over cosmic time, and its relation to galaxy mass and star-formation. Furthermore, it is becoming clear that there appear to be two distinct modes of accretion of gas onto supermassive black holes. First is the efficient accretion of cold gas presumably funnelled into the central supermassive black holes after a major merger event, the so-called "quasar- mode" feedback. Second, relatively inefficient accretion of hot gas, which occurs more widely at $z < 1$, and is generally associated with the so-called "radio-mode" feedback [15]. The form of this accretion appears to be linked to the nature of the emission lines in the AGN, split into high-excitation and low-excitation emission-line galaxies, see e.g. [31][32], and thus only with spectroscopy of large samples spanning all redshifts can we measure the relative importance of these two hypothesized feedback mechanisms. The source densities expected in the 3-tier LOFAR surveys range from ~1500 per square degree in the all sky survey to 40,000 per square degree in the deepest tier. Thus wide-field MOS capability is crucial for following up LOFAR surveys, which are necessarily restricted to the northern hemisphere and thus ideally situated for the WHT.

We are now entering an era of survey astronomy at all wavelengths. With Spitzer already underway, along with the recent launch of Herschel, the mid- and far-infrared Universe is becoming accessible over huge swathes of the sky (e.g., [20]). Spectroscopic follow-up of such surveys is crucial to their scientific success. Objects discovered by Herschel will be dusty; it is well established that photometric redshifts do not work as accurately as with the unobscured population, thus spectroscopic redshifts are required. Currently the survey with the largest area being carried out by Herschel is Herschel-ATLAS, which will cover 500 square degrees, with a source density ~500 per square degree. A MOS on the WHT with a large multiplex capability would be an ideal instrument for following up the Herschel-ATLAS sources. Moreover, a new proposal is being put together to survey ~4000 square degrees with Herschel, at which point a large MOS with wide FOV on a 4-m class telescope becomes crucial.

### 3.3.2 *The timing and cause of cosmic reionization*

One of the most pressing questions in high-redshift galaxy evolution is the timing and cause of cosmic reionization. Thanks to a combination of WMAP [19] and SDSS [24] we now know that reionization occurred somewhere in the redshift interval 6<z<10. However, crucially, we still don't know whether high-redshift galaxies alone are responsible for reionization, precisely when reionization occurred or even if reionization occurred once, or several times over. Although rapid progress has recently been made on deriving the galaxy luminosity function at 6<z<8 using HST (e.g. [46]), such studies are based on very small cosmological volumes, with correspondingly large cosmic variance uncertainties. Moreover, simply knowing the number density of high-redshift galaxies will never be sufficient to understand reionization, because to make real progress we also need to measure the Lyα escape fraction. Perhaps the most promising method for establishing the Lyα escape fraction is to measure how the prevalence of strong Lyα emitters (LAEs) evolves with redshift. Using large, statistical samples spanning a wide range of intrinsic luminosities it is possible to measure the change in Lyα transmission as the IGM transforms from neutral to ionized [66]. Indeed, a simple calculation illustrates that a wide-field MOS on the WHT would be ideal for undertaking such an experiment. Recent studies have shown [52] that at 3<z<6 there are of order 1500 LAEs per square degree brighter than $L(Lyα) = 10^{43}$ erg s$^{-1}$ (observable in about 10 hours on the WHT). Therefore, a transformative survey of $\geq$ 10,000 LAEs over 10 square degrees would be feasible in roughly 20 nights on the WHT. In contrast, despite its larger aperture, VLT/VIMOS would require 100 nights to perform the same survey.

### 3.3.3 *The recent phases of galaxy cosmological evolution*

Understanding the formation and evolution of local galaxies and their dark matter halos also requires look-back studies. In the z<0.5 Universe, optimal for 4-m telescopes, large galaxies have completed the majority of their star formation but appear to still be assembling their mass, while small galaxies are still forming stars (see, e.g., [24]). Environment plays a important but poorly understood role in this evolutionary process. Clusters of galaxies at these redshifts provide ideal laboratories for the study of processes such as ram-pressure stripping, strangulation, mergers and harassment that are most common in dense regions. Velocity fields, chemical composition, stellar ages and star formation rates, all signatures and diagnostics of evolutionary processes, can be inferred from absorption and emission-line spectroscopy from both spatially resolved (integral-field) and unresolved (multi-object) spectroscopy. Current galaxy formation models envisage the build-up of the red and blue sequences [58] via environmental and mass-dependent quenching, and dry mergers, which would leave a kinematic imprint on a galaxy. A survey targeting galaxy formation and evolution out to moderate redshifts will be composed of a number of observational probes, including studies of the Tully-Fisher relation in intermediate-redshift clusters, to measure luminosity evolution in disks from z = 0.5 to the present [3][48]; measurement of the internal dynamics of early-type dwarf galaxies, to investigate the effect of environmental and internal processes upon their scaling relations [28][72][38]; measuring the degrees of rotational and pressure support in early-type galaxies, and their environmental dependences [67]; determinations of the ages and metal abundances of the stellar populations in galaxies in different environments, to investigate the timescales for quenching star formation and how they depend upon mass and environment [64][65]; direct measurement of the dynamics of the build up of clusters and groups of galaxies, through infall along filaments at z ~ 0.1-0.3 [4]; and measurement of the evolution of gas content and star formation rates, and the origin of the Butcher-Oemler effect [13].

A single IFU and multiple mini-IFU feeds are ideal solutions for many of these studies. The most demanding ones seek the measurement of galaxy internal velocity dispersions of 20 - 30 km s$^{-1}$, which requires resolving powers of 5000 – 7000. A 2 degree field is required for cluster studies at z ~ 0.03, where this corresponds approximately to twice the virial radius. For more distant clusters and groups, a wide-field monolithic IFU (Sec. 7) such as the one proposed for nearby galaxy disks (Sec. 3.3.4) provides an efficient input mode.

### 3.3.4 *Total and baryonic masses of galaxies*

The masses of observed galaxies are a critical consideration for testing galaxy mass-assembly histories predicted within the current ΛCDM paradigm of structure formation and the baryonic physics that govern a galaxy's visible content. Indeed, fundamental scaling relations between the observable properties of galaxies (such as the Tully-Fisher, luminosity-metallicity, and Fundamental Plane relations) often tighten when one converts light to mass based on a priori knowledge or assumptions concerning the link between a galaxy's spectral energy distribution (SED) and its baryonic mass content [10][48][68]. Therefore, understanding these relations and their physical implications depends strongly on reliable mass measurements and, in fact, the differentiation between baryonic and dark-matter (DM) mass. The conventional approach to measuring the baryonic masses of stellar-dominated systems is based on the (often gross) reproduction of the observed SED via stellar population synthesis models; however, such an approach rests on the shaky foundations of an assumed initial mass function for the stellar population as well as on star-formation and chemical-enrichment histories.

The measurement of total masses requires kinematic measurements. Nearby spiral galaxies are ideal for these studies. Classical kinematic tracers have been ionized gas [56] and neutral hydrogen [70]. In the rotation-dominated regime of disk galaxies, total masses within a given radius can be determined by measuring the rotation curve; however, the decomposition of the baryonic and DM mass contributions depends critically on the adopted mass-to-light ratio (M/L) of the disk component. An oft-used means of circumventing direct measurements of M/L is the maximum-disk hypothesis [69]; however, this hypothesis remains unproven. In fact, the DiskMass Survey [8][9], using direct measures of the disk-mass surface densities of nearly face-on galaxy disks via the out-of-plane motions of disk stars, finds the maximum-disk hypothesis often over-predicts the masses of galaxy disks by as much as a factor of 2. This has serious implications for our current understanding of the mass budget of the universe at high redshift. However, the DiskMass Survey is effectively a diameter-limited sample of only ~40 galaxies and is, therefore, somewhat limited in its statistical relevance. The larger FOV of the large wide-field IFU (LIFU) described here, as compared to the SparsePak and PPak IFUs, will provide for a much-needed extension of this concept to (1) larger sample sizes by including galaxies of greater angular scale and (2) greater sensitivity obtained by a larger number of fibers, a more efficient spectrograph, and broader simultaneous spectral coverage. This extension will provide for greater dynamical range in global galaxy properties, including the ability to probe the crucial low-surface-brightness regime – posited to be different from the high-surface-brightness regime, as such galaxies should be DM-dominated at all radii.

Measurements of disk-mass surface densities following the procedure used by the DiskMass Survey require a decomposition of the line-of-sight (LOS) velocity dispersions into a stellar velocity ellipsoid (SVE) as a function of galaxy radius. In the same way as follow-up measurements to GAIA discussed in sections 2.1 and 2.2 reveal the full phase-space distribution function (DF) of the Milky Way, measurements of the SVE yields the DF in external galaxies for an ensemble average of their disk stars. Thus, measurements of the SVE in external galaxies and the full DF in the Milky Way are mutually beneficial: the former provide a cosmological context for the latter and the latter can lend physical constraints on allowed solutions to the former. Moreover, the measurement of the SVE in external galaxies lends many pertinent dynamical insights into galaxy disks in addition to the disk-mass surface density. In particular, the integral-field description of the SVE quantifies secular disk heating mechanisms, disk stability, the influence of streaming motions due to density waves, and the shape of the disk potential. Such measurements will provide novel insights into the evolution and self-regulation of galaxy disks.

Measuring LOS velocity dispersions down to 10 km/s in the outer parts of spiral-galaxy disks requires spectroscopy with a resolution of R=10,000 at surface brightness levels reaching $\mu_B$=24.5 mag/arcsec$^2$ –the typical value at three disk scale lengths– and beyond. A high-grasp, wide-field (~2') integral-field unit can collect far more light than conventional, e.g. long-slit, spectroscopy and provides 2D coverage in a single pointing, which is critical for this science. Moreover, IFUs with contiguous spatial coverage allow for fibre averaging when necessary to increase signal-to-noise. This yields clean measurements well beyond two disk scale-lengths, where contamination from bulge stars is negligible and the rotation curve of the stellar disk has reached its flat part. At the same time, innovative hybrid fibre packing designs — a core of small-diameter fibres surrounded by a halo of large-diameter fibres — permits the study of the galaxy centers at higher spatial (and spectral) resolution, which is particularly relevant for studies of low-surface-brightness galaxies.

## 4  TOP-LEVEL SCIENCE REQUIREMENTS

The science cases discussed in the previous section each leads to specific science requirements. Two alternatives present themselves: to build a separate instrument for each set of requirements, or, alternatively, to explore what compromises may allow accommodating more science, taking into account complexity and cost implications. In this section we outline a set of science requirements that produce competitive capabilities to address the science case. These are being derived from discussions among interested members from the ING community, also taking into account recommendations from the ASTRONET ETSRC report.

### 4.1  Front ends

The instrument should have three front ends: (1) a massively multiplex MOS mode for point sources, with fibres that project to about 1.2 arcsec on the sky; (2) a large integral field unit (LIFU) with a FOV above 1 arcmin in diameter, with wide (2-3 arcsec) fibres; (3) a third front end featuring multiple deployable mini-IFUs.

## 4.2 Field of view

The FOV of the MOS mode should be as large as feasible on the WHT, in order to increase survey speed for target samples with moderate densities, such as bright (V>15) halo and thick disk stars (few hundred per square degree). Current designs of an optical corrector for the WHT PF show that the largest unvignetted FOV will be 2 degrees in diameter, i.e., an area of π square degrees.

## 4.3 Spectral resolution

The science cases presented in Sec. 3 call for a range of spectral resolving powers. At the low end, cosmology redshift surveys can work with R=1,000. Radial velocity work linked to Gaia, as well as many galaxy evolution programs, work well with R~5,000. The core science of the LIFU system needs R~10,000, while chemical abundance determinations require R~20,000. We set the spectral resolving power requirements for the instrument to be R=5,000 and R=20,000. A resolution of R=5,000 fulfills the requirements for Gaia radial velocities and galaxy evolution programs; redshift surveys for cosmology can also efficiently work with R=5,000, particularly with on-chip binning capability. These resolutions are understood for nominal 1.2 arcsec diameter fibres; with the thicker fibres of the LIFU, the high-resolution mode will yield a true resolving power close to the requirement of R=10,000 for galaxy disk studies.

As discussed in Sec. 10, the R=20,000 may not be required if, for instance, such mode becomes available independently on another telescope. If this happened, we would retain the requirement for R=10,000 resolution.

## 4.4 Wavelength range

In the R=5,000 mode the minimum coverage should be from 480nm to 900nm in one setting, in order to provide the full velocity information in the spectrum from the Mg region at 517nm to the 855nm Ca triplet region. If feasible, a broader coverage from 370nm ([OII]) to 980nm in one setting should be sought, to accommodate both nebular emission-line objects in the Milky Way on the blue side, and redshifted optical emission lines for cosmology redshift surveys. While CCDs are sensitive redward of this limit, data becomes progressively intractable due to the abundance of telluric lines hence there is little gain in extending the wavelength coverage to, say 1100nm.

At resolution R=20,000, spectra should cover 200nm, from 480nm to 680nm, in a single setting.

## 4.5 Multiplex

The required multiplex depends on both the target density and the required number of targets. For Gaia follow-up radial-velocity and abundance measurements, multiplexes of 1000 and 200-300 respectively are required. For galaxy-evolution surveys, a multiplex of a few hundred is required. For cosmology redshift surveys, multiplex needs to be higher than 1000; values of 2000 or 3000 lead to reasonable survey times.

## 4.6 Throughput

Notwithstanding the survey speed advantages of our WHT WF MOS when compared to a MOS on a 10-m telescope, all of the projects presented in Section 3 are photon-starved. Throughput needs to be maximized in order to increase the competitive edge of the instrument.

## 4.7 Exposure times

This requirement puts constraints on the field configuration time, given the basic requirement that surveys should make efficient use of telescope time. Current estimates suggest that the majority of exposure times will be ~1 hour for MW and cosmology programs; 4 hours for most galaxy-evolution surveys, and up to 20 hours in a few cases.

## 4.8 Survey duration

The envisoned surveys require a few thousand nights for cosmology redshift surveys, 600 to 800 nights for MW projects, and several tens of nights for galaxy evolution surveys. Assuming that nights can be shared among various projects, the planned MOS science may require of the order of 2000 WHT nights, or 7 years. However, with ~2000—3000 fibres at R=5000 it is likely that both the Gaia and cosmology science cases can be carried out with a single commensal survey.

# 5 INSTRUMENT CONCEPT

The requirements listed in Sect. 4 can be met by a spectrograph housed in one of the WHT's Nasmyth enclosures, and fed by up to three fibre front ends. The MOS fibres will come from the PF unit, where a FOV of 2 degree diameter will be provided by a new corrector, for which two possible designs are described by Agocs et al (2010, this volume). Mini-IFU fibre units would also be fed from PF, whereas the wide-field monolithic IFU might be fed at the Nasmyth focus. The scheme resembles the current successful setup at WHT, where the WYFFOS spectrograph on the GHRIL Nasmyth platform†† is fed either with MOS fibres from the AF2 fibre positioner at PF, or from the INTEGRAL IFU at Nasmyth (AF2 and WYFFOS handle up to 140 fibres on a FOV of 40 arcmin diameter, hence they insufficient for the survey capabilities discussed here).

# 6 MULTI-OBJECT FRONT END

The history of fibre multi-object spectroscopy includes many different concepts for the deployment of fibre pick-offs within the telescope focal plane [63]. Broadly speaking, these can be subdivided into four main types in order of complexity: Plug plates (FOCAP [30], SDSS [53]), fishing poles (MX [34]), pick and place (AUTOFIB [54] AF2 [42], 2dF [43], HectoSpec [23], FLAMES [27]) and locally sampling (FMOS [29], SIDE Concept [59], WFMOS Concepts [50][26]). For each approach there are advantages and disadvantages, as broadly outlined in Table 1.

Table 1. Fibre pick-off options: advantages and disadvantages.

| Positioner Concept | Advantages | Disadvantages |
| --- | --- | --- |
| Plug Plates | High Density, low complexity, low build cost, close target proximity, curved focal surfaces. | Handling, offline machinery requirements, labor intensive – high operations cost. |
| Fishing poles | Independent units, modular system. | Low multiplex, poor crowding/proximity constraints. |
| Pick and place | Flexible, close target proximity. | Single point failure mode. |
| Locally sampling | Scalable, robust, high density, uniform areal sampling, curved focal surfaces acceptable. | Complexity, high cost, impact of clustering. |

In the context of the current study we aim to investigate a number of possible solutions that could deliver the required specifications. The final choice for the positioner will be driven by our goal of maximizing the multiplex of the system while staying within realistic cost and mass envelopes. We will pay specific attention to solutions that can be implemented efficiently in terms of the cost and staff effort required for both implementation and support. This approach naturally leads to the investigation of the extent to which off-the-shelf components can be used, and the extent to which present-day commercial positioning systems can meet the demanding specifications required for the prime-focus environment. For a 1.5 arcsec fibre at the 57μm/arcsec plate scale of the WHT prime focus, this implies a positioning tolerance of ~9μm, compared to the 20μm specification for the 2dF pick and place positioner, or 10μm for the FMOS Echidna positioner [1].

In terms of the multiplex requirements highlighted in Section 4.5, we note that the 2dF instrument deploys 400 fibres within a similar field size to that proposed here (~500mm diameter). This number was close to the limit for the combination of fibre button and retractor designs used by 2dF. We have revisited the design concept of 2dF and find that, twenty years on, off-the-shelf systems can be expected to deliver 5μm positioning accuracy with high speed and acceptable mass, and therefore they may provide a solution for a new positioner for the WHT. This scheme would require an extension of the pick-and-place concept to a 3-dimensional configuration with multiple retractor heights, to reduce the limitation of attempting to pack all the fibres into a single circle at the field edge. Modeling this concept with workable fibre buttons (the same size as the 2dF buttons) shows that ~1000 fibres can be usefully deployed within a 2 degree FOV (Figure 1). Notably, as can be seen in the figure, multiple retractor heights allow for fibre crossing. This concept shows that we can meet the requirements for the Gaia follow-up science, with some scope to incorporate deployable IFUs within the same pick-and-place framework.

---

†† http://www.ing.iac.es/PR/wht_info/whtghril.html

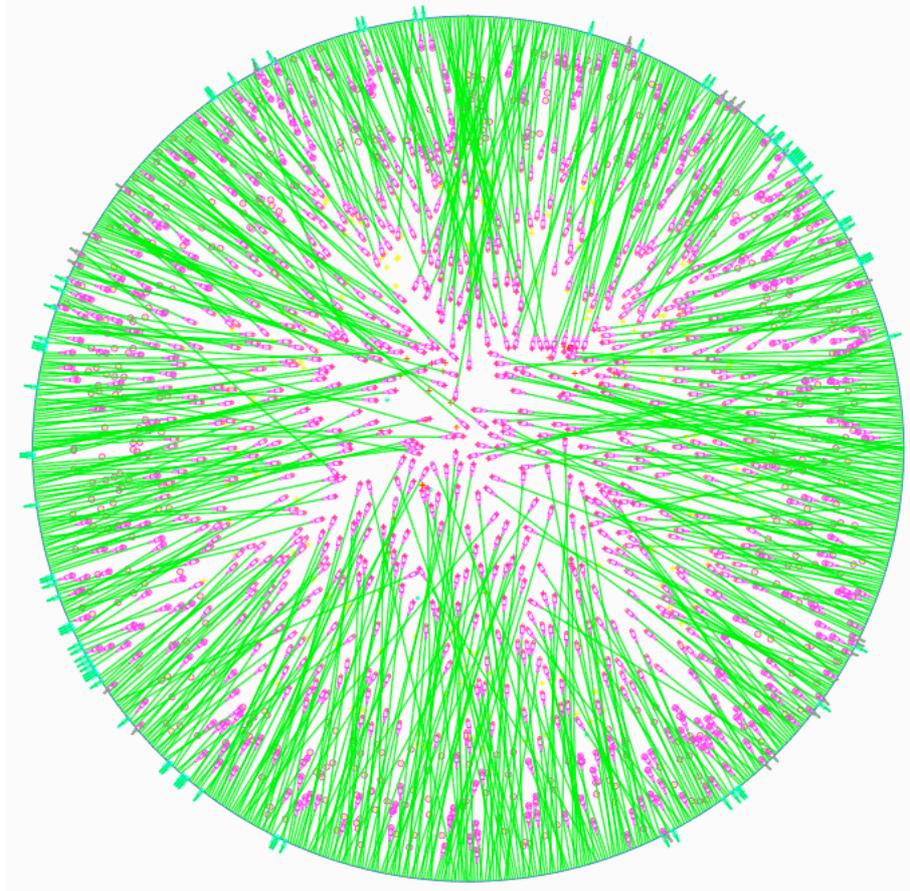

Figure 1. Sample configured field with 1000 2dF buttons on a 2 degree diameter focal plane. 929 fibres have been successfully deployed on a random distribution of targets in this example.

Meeting the objective of >1000 target fibres for the cosmological surveys will be extremely challenging for this concept, and so will require a revisitation of Echidna- [1] or Cobra-like [26], designs. As an example, the FMOS Echidna successfully deploys 400 fibres within a physical field of view of 150mm diameter. Scaling an Echidna-like concept up to the WHT FOV would allow for up to 3000 fibres to be accommodated. But these solutions have specific challenges. The multiplex of 3000 actually is a requirement of an Echidna implementation for our MOS, as, when covering the entire focal plan with less fibres, the non-telecentricity losses as you move a fibre off-axis become unacceptable. The mass of a Echidna-based system is expected to be of order 1000kg, compared to ~400kg for a pick-and-place system. Cost is also expected to be notably higher. Finally, a focal plane populated with Echidnas makes the deployment of mini-IFUs difficult; we would require enough back-focal distance from the corrector to deploy MX-type arms or a simpler pick-and-place solution at the focal plane for the mini-IFU.

# 7 WIDE-FIELD LIFU FRONT END

Measuring stellar velocity dispersions in the kinematically cold outskirts of galaxies requires a spectral resolution of R=10,000 for extended emission at low surface brightness levels, typically 2-3 magnitudes below the sky brightness. Maximizing the signal-to-noise in the spectra requires the highest achievable spectrograph efficiency and a maximum specific grasp for the fibers to collect the largest amount of light per fibre. The distribution of stellar light in the outskirts of galaxies is smooth and achieving spatial resolution is not a requirement. Therefore, fibre sizes should be driven to the largest possible diameter that can still yield a spectral resolution of R=10,000. The signal-to-noise ratio can further be improved by averaging the spectra from fibres that are adjacent on the focal plane, sampling the same physical region of the galaxy. A 100% filling factor of the mono-IFU is not required and can, in any case, be achieved by dithering offsets.

With spatial resolution of little concern, the specifications for the monolithic LIFU are reasoned here 'from the slit to the focal plane' under the assumption that the spectrograph has a high-resolution mode. Assuming that 90μm fibre cores yield R=20,000, the maximum allowed fibre diameter follows from the optical layout of the spectrograph. In the high-resolution mode, R=10,000 can still be achieved with fibres that have a core diameter of 180μm. The available slit-length of 230mm can accommodate 1045 such fibres, assuming an outer fibre diameter of 220μm and allowing for sufficient distance between the fibres to minimize cross-talk between the spectra on the detector. To maximize the specific grasp of the fibres, they can be fed at f/3.6 in the Nasmyth port by employing a focal reducer lens to speed up the incoming beam from f/11. In the focal plane, 919 fibres can be packed in a hexagon with a hexagonal grid and a central row of 35 fibres. This allows for another 120 fibres to be positioned on the sky. A central row of 35 fibres with an outer diameter of 220 μm, packed buffer-to-buffer, spans 7.7mm in the focal plane. This corresponds to a field-of-view of the IFU of 1.985 arcmin and a fibre core diameter of 2.6 arcsec.

This LIFU design, albeit with coarse spatial resolution, is also suitable for 3D spectroscopy of distant groups and clusters, when used with the R=5000 mode of the spectrograph.

# 8 MINI-IFU FRONT END

An important mode for the spectrograph is the multiple, deployable mini-IFU mode, which allows the analysis of velocity fields of samples of galaxies and other extended objects. Multi-IFUs are a comparatively new development; the first in common use is that for the GIRAFFE spectrograph on the ESO VLT UT2 [27]. This offers 15 IFU units of 20 spatial pixels each. Mini-IFU feeds are being developed for the ESO ELT (the OPTIMOS-EVE and EAGLE projects), and in the infra-red for KMOS on the ESO VLT [62].

There is a clear tradeoff between the size and number of IFU units given the boundaries imposed by the slit length of the spectrograph. For many of our science drivers a velocity field is required, and a minimum of 100 spatial pixels per IFU is required. In Table 2 we give a baseline set of requirements for a single spectrograph, and a second, optimal, set assuming that we can provide three identical copies of the spectrograph.

Table 2. Requirements for the mini-IFU functionality

| Parameter | Baseline spec | Optimal Spec |
|---|---|---|
| Resolving power | 5000 | 7000 |
| Wavelength range | 400-950nm | 370-980nm |
| Number of IFUs | 10 | 21 |
| Spatial pixels per IFU | 10x10 | 14x10 |
| Spatial Pixel size | 1 arcsec | 1 arcsec |
| IFU footprint (arcsec$^2$) | 10x10 | 14x10 |
| Field of view | 2 degrees | 2 degrees |
| Number of fibres | 1000 | 2940 |

# 9 SPECTROGRAPH

The goal is to study the ability of various spectrograph designs to accommodate the science requirements. Here we present first ideas of what is very much work in progress, largely focussed on the R=5,000 mode.

Our base-line concept is a dual-beam design delivering R=5,000 resolution with 1.2 arcsec fibres. It assumes that the output from the telescope PF corrector will be lensed to provide an input beam at ~f/3.6 (an input f/2.8 beam as delivered by the corrector would require a costly ~f/2.7 collimator). A consideration of different possibilities for the focal plane format with this configuration gives a set of basic parameters outlined in Table 3. We have considered 8kx8k and 6kx6k camera formats (15 μm pixels) that allow for optimal sampling of the full spectral range with two cameras. For this study we fix the input fibre core at 90 μm and allow some flexibility of the fibre-fibre spacing at the slit to illustrate how cross-talk considerations affect the multiplex. For illustration, we have also considered the accommodation of a 2 arcsec fibre core LIFU bundle to determine the highest spectral resolution that could reasonably be achieved with this design. Here we note that a larger fibre core relaxes the image quality requirement for the camera for a given spectral resolution, moving the trade from image quality to spectral coverage.

We have investigated the possibility to push the spectrograph performance up to R~30,000 by means of a grating change. This appears possible in theory, but is likely to lead to sub-optimal sampling of the spectral resolution elements, and so the question of whether/how to achieve the high-resolution mode remains open at this stage.

Table 3. Some possible spectrograph parameters for the R~5,000 mode. Higher multiplex than the numbers in the right column would be implemented by increasing the number of spectrographs.

| Detector format | Camera f/ratio | Input core @ f/3.5 | Blue range (nm) | Red range (nm) | No. of fibres |
|---|---|---|---|---|---|
| 6kx6k | 1.75 | 90μm, 1.25 spacing | 377-529 | 531-764 | 920 |
| 8kx8k | 1.75 | 90μm, 1.75 spacing | 366-597 | 585-953 | 990 |
| 8kx8k | 1.75 | 150μm (LIFU) | 440-520 (R=16000) | 720-850 (R=16000) | 700 |
| 6kx6k | 1.4 | 90μm, 1.75 spacing | 369-609 | 607-1004 | 930 |

The numbers in Table 3 are derived from top-level considerations of the spectrograph parameters. Figure 2 shows a preliminary design for the blue arm of the f/1.4 camera option shown in the last row of Table 3, and provides some more detailed information on the spectrograph space envelope and general feasibility.

This design accommodates a slit length of 230mm for a collimator field of view of ±9.3°. The collimated beam size is 200mm, with an 18.5° grating angle, which gives a grating format that can be fabricated using existing facilities. The 90 micron fibre core projects to 2.4 pixels (15 micron pixels) on the detector, and the fibre-fibre spacing at the slit gives 3.75 pixels centre-centre between adjacent fibres. These numbers imply that the camera image quality should be ~2pixels rms spot size or better. The largest camera lens is 220mm diameter, which should be achievable, although the design needs to be controlled to ensure that the required blank sizes for all lenses remain feasible. The corresponding red camera is shown in Figure 3. Throughput estimates, for each component and for the total system including telescope, are given in Figure 4.

This design lends itself to delivering higher spectral resolutions by changing the grating angle and placing the camera on an articulated arm. A resolving power of R~10,000 can easily be reached this way. In principle we could push the angle to yield an R=25,000 output with a 55° grating angle and a set of four higher density gratings required in each arm to achieve full spectral coverage, but the camera image quality must increase to better than 1 pixel rms spot size for this to be practical.

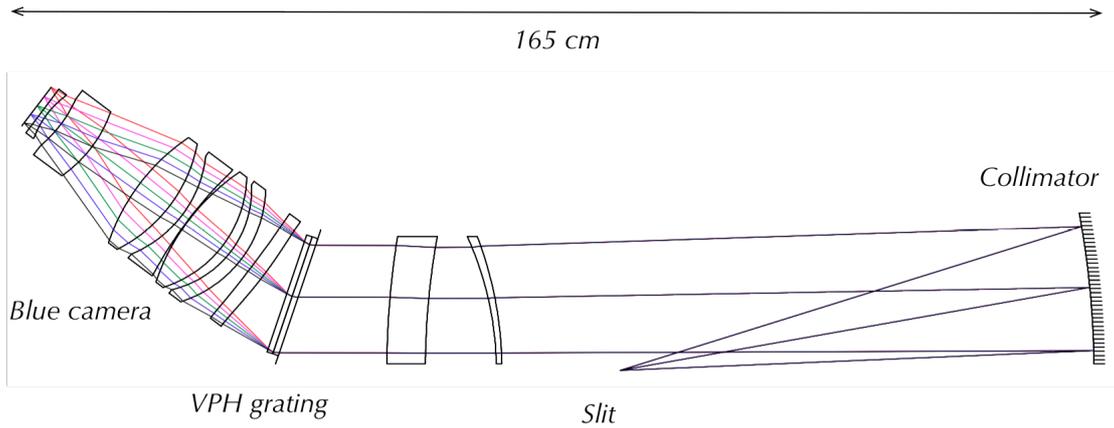

Figure 2. Preliminary design for the blue arm. The collimated beam of 200 mm is dispersed by a VPH grating and focused by an f/1.4 camera.

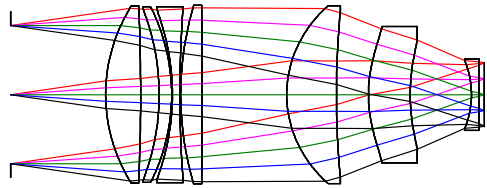

Figure 3. Red-arm camera design.

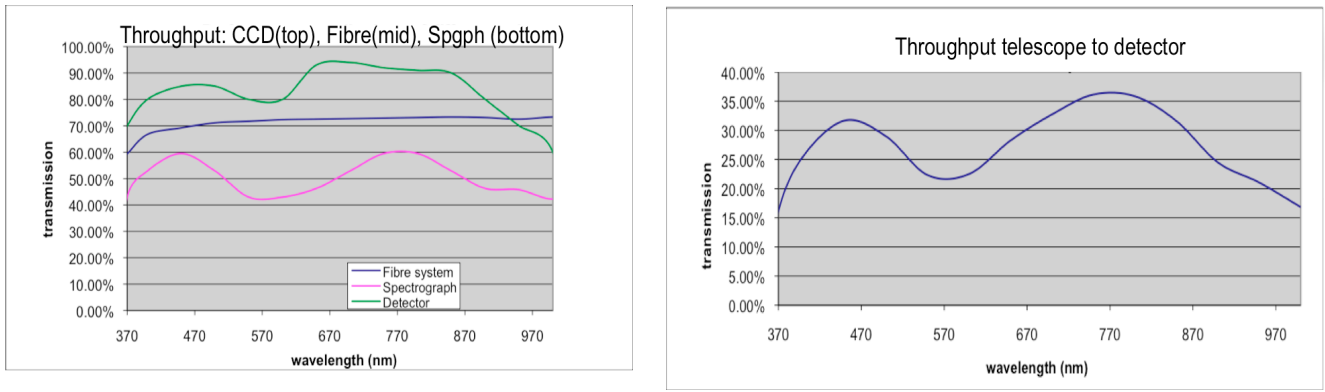

Figure 4. Throughput estimates for the R=5,000 mode of the WHT MOS. *Left:* top to bottom, the contributions of detector, fibre and spectrograph. *Right:* total throughput from the telescope to the detector.

## 10   MAIN TRADE-OFFS

The science objectives of a MOS on the WHT demand, above all, a wide FOV: for most surveys envisioned in the science cases, the access to a larger number of sources in a single exposure directly translates into faster survey speed, hence realizing the multiplex advantage of the WHT vs 10-m telescopes.

For a given FOV (2 degrees as currently planned for the future WHT PF corrector), the required multiplex varies strongly from one science case to another, and scales with the surface density of sources that one may want to observe at the same time, taking into consideration survey planning factors such as the advantage of obtaining similar signal-to-noise for most sources in a given exposure. For redshift surveys for cosmology, the multiplex should be as high as possible, and this also facilitates planning of surveys in which it may be desirable to simultaneously observe several types of science targets. Having a spectrograph that accommodates a high number of fibres benefits not only the MOS mode but also allows the LIFU and mini-IFU modes to deploy IFUs with a larger number of fibres.

Hence, multiplex will likely be limited by technology and by aspects such as weight and cost. We understand that the baseline multiplex is $1000/\pi$ per square degree, but that we should explore the cost implications of a higher multiplex.

Weight, cost, and configuration times strongly affect the positioner choice. Pick-and-place robots are attractive from this point of view, and their technology is well understood. However, packaging constraints limit the multiplex to up to ~1000. At the other end in the complexity scale, locally-sampling Echidna- or Cobra-type systems may have their multiplicity limited by weight and cost constraints; they also make the deployment of mini IFUs difficult. The vast majority of the exposures envisaged in the currently planned surveys will be ~1 hr. Efficient use of the telescope time then requires field changeover times below ~5 min. For a locally sampling positioner this becomes a direct time constraint, whereas for a buffered pick and place positioner, the field configuration time should be < 1 hour.

Weight and cost strongly affect the choice of the type of positioner. Pick-and-place robots are attractive from this point of view, and their technology is well understood. However, configuration times probably limit the multiplex to values up to ~1000. At the other end in the complexity scale, systems may have their multiplicity limited in the end by weight constraints, and are more expensive. Configuration times affect the choice of the positioner. Perhaps surprisingly, the vast majority of the exposures envisaged in the currently planned surveys will be of order 1 hr. Efficient use of the telescope time then requires configuration times below say 20 min. Locally-sampling concepts (Table 1) have the potential to strongly out-perform robot positioners in that regard.

Planning the provision of three distinct input modes (MOS, LIFU, mini-IFU) does not strongly constrain the design of the MOS. Each mode simply translates into a separate front end for the same spectrograph, and opens, for little cost, an entirely new science mode. Risks associated to the provision of different inputs are mostly related with the fact that each mode likely requires a separate specific data acquisition strategy and data reduction pipeline. The LIFU mode does add specific complexity to the spectrograph, as it requires spectral resolutions higher than the reference resolution R=5000. If the spectrograph does accommodate a high R mode, then the impact of the LIFU on the spectrograph complexity is low. If the final choice is for the exclusion of the R=20,000 mode, then adding the LIFU mode does increase the complexity of the spectrograph.

Whether the MOS should provide for a high-R mode, in addition to R=5000, is among the most critical decisions for this instrument. The spectrograph without the high-R mode will be simpler, smaller and cheaper; and a smaller size might in turn facilitate higher multiplex, since more replicas of the spectrograph would fit inside the GHRIL room. However, the high-R mode is critical to the science exploitation of the Gaia data and the study of the dynamics of the disks of nearby galaxies. Considerations such as the existence of another high-R MOS on a telescope to which the community has access, and the sources of funding, may ultimately decide whether the WHT MOS should include a high-R mode.

A trade-off probably exists as well between resolution and throughput. In that regard, including a high-efficiency R=1000 mode may prove beneficial for cosmology redshift surveys.

## 11  CONCLUSIONS

A 4-m telescope equipped with a large field of view and a highly multiplexed multi-object spectrograph has the potential to out-perform a 10-m telescope. In this paper we have presented the key science drivers for building such an instrument. The implied science requirements are: FOV above 2 deg, multiplexing of or above 1000, wavelength range 370 - 980 nm, and spectral resolution R at least 5000. R = 5000 delivers much of the science (Gaia radial velocity follow-up, cosmology, galaxy evolution). R = 20,000 is required for stellar abundance determinations (Gaia follow-up), and R = 10000 for niche work on galaxy disks.

A multiplicity of front ends (MOS, IFUs, mini-IFUs) is an attractive option, providing considerable extra functionality for little additional cost.

Initial modelling suggests that a double-beam spectrograph with VPH dispersers would meet the science requirements for the R = 5,000 mode.